%% file: main.tex
\documentclass[10pt,conference]{IEEEtran}
\IEEEoverridecommandlockouts
\usepackage{cite}
\usepackage{amsmath,amssymb,amsfonts}
\usepackage{algorithmic}
\usepackage{graphicx}
\usepackage{tabularx,booktabs}
\usepackage{textcomp}
\usepackage{xcolor}
\usepackage{url}
\usepackage{multirow}
\usepackage{enumitem}
\usepackage{color, colortbl}

\definecolor{codegreen}{rgb}{0,0.6,0}
\definecolor{codegray}{rgb}{0.5,0.5,0.5}
\definecolor{codepurple}{rgb}{0.58,0,0.82}
\definecolor{backcolour}{rgb}{0.95,0.95,0.92}
\definecolor{codekeyword}{RGB}{153, 51, 153}

\usepackage{listings}
\lstdefinestyle{mystyle}{
  backgroundcolor=\color{white},   commentstyle=\bfseries\color{codegreen},
  keywordstyle=\bfseries\color{codekeyword},
  numberstyle=\tiny\color{codegray},
  stringstyle=\color{codepurple},
  basicstyle=\ttfamily\scriptsize\linespread{1.15},
  breakatwhitespace=false,         
  breaklines=true,                 
  captionpos=b,                    
  keepspaces=true,                 
  numbers=left,                    
  numbersep=5pt,                  
  showspaces=false,                
  showstringspaces=false,
  showtabs=false,                  
  tabsize=2,
  xleftmargin=16pt,
  xrightmargin=0pt,
  breakindent=0pt,
  resetmargins=true
}
\usepackage[framemethod=TikZ]{mdframed}
\mdfdefinestyle{style1}{
innerleftmargin=0.4cm,innerrightmargin=0.4cm,
innertopmargin=0.3cm ,innerbottommargin=0.3cm,
roundcorner=10pt,linewidth=1.0pt,
footnoteinside=false}

\def\BibTeX{{\rm B\kern-.05em{\sc i\kern-.025em b}\kern-.08em
    T\kern-.1667em\lower.7ex\hbox{E}\kern-.125emX}}
    
\newcolumntype{Y}{>{\centering\arraybackslash}X}

\begin{document}

\title{Using mutation testing to measure behavioural test diversity \thanks{We acknowledge the support of Vetenskapsr\aa det (Swedish Science Council) in the project BaseIT (2015-04913).}
}

\author{\IEEEauthorblockN{Francisco Gomes de Oliveira Neto, Felix Dobslaw, Robert Feldt}
\IEEEauthorblockA{\textit{Chalmers and the University of Gothenburg} \\
\textit{Dept.\ of Computer Science and Engineering}\\
Gothenburg, Sweden \\
francisco.gomes@cse.gu.se, \{dobslaw,robert.feldt\}@chalmers.se}
}

\maketitle
\thispagestyle{plain}
\pagestyle{plain}

\begin{abstract}
Diversity has been proposed as a key criterion to improve testing effectiveness and efficiency. 
It can be used to optimise large test repositories but also to visualise test maintenance issues and raise practitioners' awareness about waste in test artefacts and processes. 
Even though these diversity-based testing techniques aim to exercise diverse behavior in the system under test (SUT), the diversity has mainly been measured on and between artefacts (e.g., inputs, outputs or test scripts).
Here, we introduce a family of measures to capture behavioural diversity (b-div) of test cases by comparing their executions and failure outcomes.
Using failure information to capture the SUT behaviour has been shown to improve effectiveness of history-based test prioritisation approaches. 
However, history-based techniques require reliable test execution logs which are often not available or can be difficult to obtain due to flaky tests, scarcity of test executions, etc.
To be generally applicable we instead propose to use mutation testing to measure behavioral diversity by running the set of test cases on various mutated versions of the SUT.
Concretely, we propose two specific b-div measures (based on accuracy and Matthew's correlation coefficient, respectively) and compare them with artefact-based diversity (a-div) for prioritising the test suites of 6 different open-source projects.
Our results show that our b-div measures outperform a-div and random selection in all of the studied projects. The improvement is substantial with an average increase in average percentage of faults detected (APFD) of between 19\% to 31\% depending on the size of the subset of prioritised tests.
\end{abstract}


\section{Introduction}
Testing is an essential activity that drives and ensures quality of software-intensive system. However, the increasing growth and complexity of software systems in combination with limited resources creates prohibitive testing scenarios~\cite{Memon2017_google, deOliveiraNeto2018_astCI}. Test optimisation approaches, such as test prioritisation or selection, cover a variety of strategies and technique that aim for cost-effective testing by automatically or manually supporting testers in deciding what and how much to test~\cite{Yoo2012}. Particularly, diversity-based techniques perform among the best when comparing a large set of test prioritization techniques~\cite{Henard2016}, hence being a prominent candidate for cost-effective testing in prohibitive situations.

Test diversity can measure how different tests are from each other, based on distance functions for each pair of tests~\cite{Feldt2008,Cartaxo2011,Hemmati2013}, or even over the entire set of tests altogether~\cite{Feldt2016}. In other words, the goal is then, to obtain a test suite able to exercise distinct parts of the SUT to increase the fault detection rate~\cite{Feldt2008,Feldt2016}. However, the main focus in diversity-based testing has been on calculating diversity on, typically static, artefacts while there has been little focus on considering the outcomes of existing test cases, i.e., whether similar tests fail together when executed against the same SUT. Many studies report that using the outcomes of test cases has proven to be an effective criteria to prioritise tests, but there are many challenges with such approaches~\cite{Kim2002_failrate,Zhu2018_cofail}. Particularly, gathering historical data to evaluate test failure rates has its own trade-off. On one hand, historical data can be readily mined or provided by many build and continuous integration environments. On the other hand, historical data can be unreliable due to, e.g., many flaky test executions, or insufficient\slash limited in case the project is not mature enough or many test cases are being added or removed from test repositories~\cite{Zhu2018_cofail,Haghighatkhah2018_jss}. Moreover, historical data is inherently connected to the quality of the test suite, such that test cases that rarely fail can hide many false negatives (i.e., tests that pass but should fail)~\cite{Memon2017_google,Zhu2018_cofail}. In short, calculating diversity based on historic test outcomes has shown a lot of promise but is limited to situations when reliable and extensive test logs are available.

As an alternative, we here \textit{propose to use mutation testing to overcome these challenges and enable more general use of diversity approaches for test optimisation}. An added benefit of our approach is that it allows controlled experiments on the value of test diversity itself. By introducing many mutants in the code and running the entire test suite for each mutant, we identify patterns in the test outcomes (passes and failures) of pairs of test cases. We call this behavior-based diversity (b-div) since it measures diversity in the behavior of the test cases when exercising different faults in the SUT. Unlike artefact-based diversity (a-div) measures, our approach considers test outcomes independently from the test specification (e.g., requirements, test input, or system output data). In other words, our distance measures conveys whether tests have been passing\slash failing together when executed on the same instance of the SUT, i.e., when it was exposed to a certain (set of) mutant(s). Note that other techniques use co-failure, to obtain the \textit{probability distribution} of an arbitrary test B failing given that test A failed~\cite{Zhu2018_cofail}, whereas our approach measures the \textit{distance} value between tests A and B based on whether they have the same test outcome.

Our goal in this paper is to describe and then investigate the benefits and drawbacks of using behaviour-based diversity based on mutation testing. Our empirical study applies the approach for test prioritisation which is one of the more direct use cases for test optimisation~\cite{Yoo2012}. We also contrast different diversity-based techniques in terms of their fault detection capabilities. The research questions we study are:

\begin{mdframed}[style=style1]
\begin{description}[align=left]
    \item[RQ1:] How can we represent and capture behavioural diversity?
    \item[RQ2:] How can we use mutation testing to calculate behavioural diversity?
    \item[RQ3:] How do artefact and behaviour-based diversity differ in terms of effectiveness?
    \item[RQ4:] What are the trade-offs in applying behavioural diversity in test prioritisation?
\end{description}
\end{mdframed}

Here, we focus on analysing open source projects and use the PIT\footnote{\url{https://pitest.org}} mutation testing tool to compare fault detection rates between behaviour-based and traditional artefact-based diversity measures. Our results show that behaviour-diversity is significantly better at killing mutants regardless of the prioritisation budget, i.e., the size of the selected test set. In particular, larger projects (containing more test cases and test execution data), yield the largest improvements. For instance, in our two larger, investigated projects, when given a budget of 30\% tests selected, the difference in the average percentage of mutants killed between the best b-div measure (MCC) and a-div measure is, respectively, 31\% and 19\%. But for tighter budgets the difference can be even larger, e.g., in one project, behavioral diversity kills 45\% more mutants when selecting only 10\% of test cases.

Here, mutants allow us to run the test against realistic faults introduced into the code, instead of measuring behavioural diversity in connection to modifications in the tests or in the SUT. Therefore, the results indicate that the use of mutation testing and behavioural diversity is a suitable candidate for prioritising tests where historical information or modifications are unavailable. In short, our contributions are:

\begin{itemize}
    \item A technique to represent and measure test diversity based on the test behaviour instead of static artefacts (e.g., test specification or code).
    \item An approach that uses mutation testing in combination with behavioural diversity to prioritise tests based on test outcomes.
    \item Two methods to measure distance using Accuracy and Matthew's correlation coefficient (MCC) between test executions.
    \item An experimental study that investigates the effectiveness and applicability of behavioural diversity for test prioritisation on open-source projects.
    \item Analysis of the differences between artefact-based and behaviour-based diversity on fault detection.
\end{itemize}

Next, we discuss the background and related work. Section~\ref{sec:technique} details our approach, along with the proposal of two distance measures to operate on test execution data through a Test Outcome Matrix (TOM). Our empirical study is described in Section~\ref{sec:methodology}, whereas the results and discussion are presented, respectively in Sections~\ref{sec:results} and~\ref{sec:discussion}. Lastly, Section~\ref{sec:conclusions} contains concluding remarks and future work.

\section{Background and Related Work}
\label{sec:background}

Generally, there are two main steps in techniques exploiting test diversity. First, we determine which information will be used to calculate diversity. Many different information sources can be used as outlined by e.g. the variability of tests model of~\cite{Feldt2008}: test setup, inputs, method calls, execution traces, outputs, etc. The second step is to use a distance function $d$ to measure the diversity of the information of each (or a subset of) test case(s). As an example, consider a test suite $T$ where $t_a,t_b \in T$ are two test cases. 

When the diversity measure is normalized, its distance function $d(t_a,t_b)$ returns a value that represents a \textit{continuum} between zero and one, where: $d(t_a,t_b) = 0$ indicates that two tests are identical, while $d(t_a,t_b) = 1$ indicate they are completely different. Numerous $d$ have been used for diversity-based testing, each supporting different types of data. For instance, the Hamming distance was applied on the traces of historical failure data based on sequences of method calls from execution traces~\cite{Noor2015}, whereas Jaccard and Levenshtein are applicable when textual information from test specifications is available~\cite{Ledru2012,deOliveiraNeto2016,Zhang2018_usecase}. Moreover, Normalized Compressed Distance (NCD) has been used as a generic distance function applicable for most types of information (e.g., full execution traces including test input and output)~\cite{Feldt2008,Feldt2016}.

Empirical studies have shown promising results when diversity is used to increase fault detection~\cite{Feldt2016,Henard2016,Haghighatkhah2018_jss}, test generation~\cite{Poulding2017_robustness} or visualization of test redundancy~\cite{deOliveiraNeto2018_apsec}. On one hand, the approaches using diversity tend to be quite general, since distance measures, such as NCD, can be used for any type of data (e.g., code, requirements, test scripts, execution traces). Particularly, when applied to test prioritisation\slash selection, diversity leverages test effectiveness by improving fault detection rate or reducing feedback cycles~\cite{Feldt2016,Henard2016,deOliveiraNeto2018_astCI}. 

On that note, there are many test prioritisation techniques proposed and evaluated in literature beyond diversity-based testing~\cite{Yoo2012}. Many prioritisation techniques focus on structural coverage, such as statements~\cite{Hao2013}, branch coverage~\cite{Jones2003}, mutants killed~\cite{Lou2015_issre}, or modified artefacts~\cite{Memon2017_google}. Other approaches use historical data, test executions and failure information to increase the rate of fault detection~\cite{Kim2002_failrate,Fazlalizadeh2009,Khalilian2012,Zhu2018_cofail,Haghighatkhah2018_jss}. In~\cite{Zhu2018_cofail} authors propose the analysis of co-failure probability of tests to drive the prioritisation, and, similar to our observations, their proposed techniques obtain significantly higher fault detection rates when compared to other approaches. However, authors measure probability values based only on tests failing together, whereas our approach measures distances between pairs of test based on their differing outcomes. Moreover, authors discuss the risks of having skewed failure data where few tests are responsible for the majority of failures. In their case, 55\% of the failures were attributed to only 4 tests. Therefore, practitioners need to verify the quality of the dataset used to calculate co-failure probabilities. For test sets with balanced failure distribution, the co-failure probability becomes a more reliable criteria for test prioritisation~\cite{Zhu2018_cofail}.

In~\cite{Haghighatkhah2018_jss}, authors investigate the use of history-based test prioritization in combination with diversity-based test prioritisation. Their approach measures distances between code artefacts, whereas failure information is used to assign weights to tests. Similar to our approach, authors identified benefits in combining both types of techniques, also leading to improved test effectiveness. However, their usage of distance measures (NCD) on large historical data introduced prohibitive time constraints, since diversity-based techniques are hard to scale due to the computational complexity of distance calculations~\cite{Cruciani2019_fastR,Feldt2016}. Our approach uses different methods to measure diversity of execution information, and the volume of data can be controlled by the mutation tools used to generate failure information, hence mitigating the scalability issues.

The approaches mentioned above consistently report improved test effectiveness when using failure information to drive prioritisation~\cite{Kim2002_failrate,Zhu2018_cofail,Haghighatkhah2018_jss}. Unfortunately, most of the issues relate to inconsistencies in the failure data, skewed failure distributions, hindrances or limitations in obtaining or operating on bulky historical data leading also to bottlenecks. Our approach uses (i) behavioral information (test outcomes\slash failures) instead of artefacts, (ii) mutants instead of historical data to obtain the failure information, and (iii) proposes different diversity measures than existing approaches. Overall this thus generalizes diversity-based testing to situations when historical data is not available or reliable, or can complement such data in scenarios when it is.

Within the mutation testing literature our approach is similar to the diversity-aware mutation adequacy criterion~\cite{shin2018theoretical}. Shin et al.,~\cite{shin2018theoretical} also describe other related mutation testing work such as that on disjoint mutants~\cite{kintis2010evaluating}, mutant (dynamic) subsumption (hierarchies)~\cite{kurtz2014mutant}, and clustering of mutants based on their similarity~\cite{ma2016mutation}. However while the adequacy criterion of~\cite{shin2018theoretical} is set-based and selects test cases that best distinguishes a set of mutations we calculate the pair-wise differences between test cases and then select the pairs with larger diversity first. Thus while their approach is a set-based maximization of the columns of a test outcome matrix (outcomes of tests by mutants) ours is a pair-based maximization of the row differences. We also quantify how different the rows are rather than just checking if the (column) vectors differ. Future work should investigate the empirical differences for fault detection that this implies, if the vector-based quantification we use here can also be extended to measure and exploit how different mutations are, and what are the trade-offs of using set-based versus pair-wise diversity maximization.

\begin{figure}
    \centering
    \includegraphics[width=0.9\columnwidth]{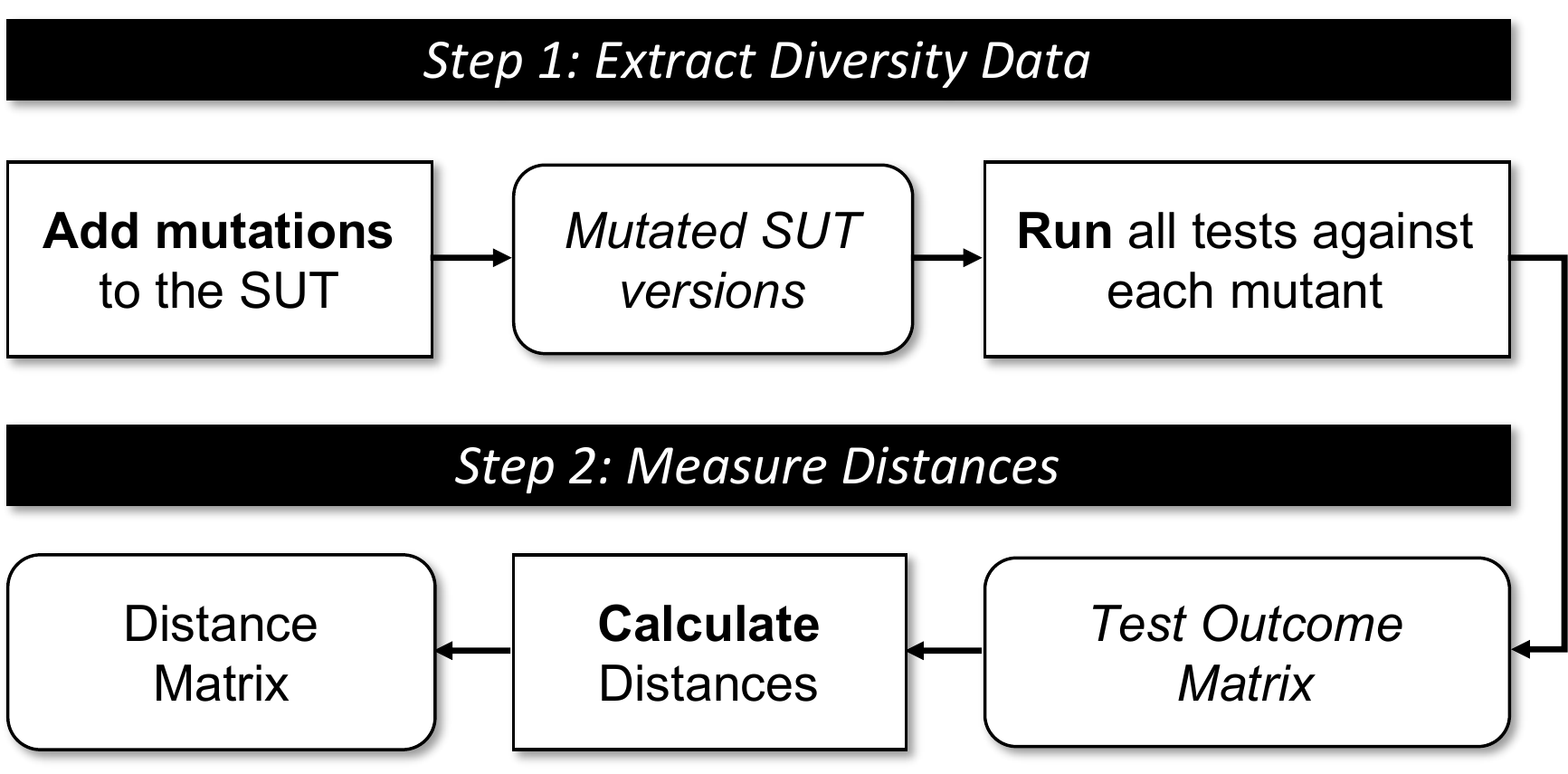}
    \caption{Overall steps of our approach describing how to extract and measure behavioural distances to test cases. The final distance matrix can then be used for test prioritization\slash selection or for other test optimisation needs.}
    \label{fig:bdiv_process}
\end{figure}

\section{Behaviour-based Diversity}
\label{sec:technique}
We describe the overall process of our approach in Figure~\ref{fig:bdiv_process}. Unlike artefact-based diversity, our technique is independent of the test specification and focuses on the test outcomes and their patterns. We collect this information in a \textit{test outcome matrix} (TOM) which includes information about the outcome (pass or fail) of different test cases (rows of the matrix) when executed for multiple mutants (columns) of the SUT.

We illustrate our approach with a toy example where our SUT (i.e., the Class Under Test) is the class \texttt{MyDate} (Listing~\ref{code:example_sut}) that, for simplicity, is a wrapper for the \texttt{LocalDate} class from Java API. Moreover, we manually created a set of corresponding JUnit tests (Listing ~\ref{code:example_test}).\footnote{\url{https://docs.oracle.com/javase/8/docs/api/java/time/LocalDate.html}}

\input{javacode_sut.tex}
\input{javacode_test.tex}

\subsection{Extracting diversity data}
We run the unit tests in five different versions of the SUT $v_{1,\cdots,5}$ where we manually introduced different mutants at the same line of code (Table~\ref{tab:toy_versions}), without changing the unit tests. We execute the tests on all five mutated versions of the SUT recording whether each test passed or fail, hence yielding the test outcomes of our different SUT versions. Since we are interested in the tests that fail together for the same fault, we run all tests against all mutants, instead of halting the execution when killing the mutant. For our technique to be applicable it is important that the mutation testing tool used allows this mode of execution. Briefly, the TOM represents the software's behaviour based on the test executions throughout the various mutated versions. Our interest is in the diversity between all rows of a TOM, such that similar rows across the different columns indicate tests with similar behaviour. In other words, our assumption is that test cases that frequently present similar outcome throughout different mutants are perceived as similar with respect to their behaviour.

\begin{table}
    \centering
    \caption{List of modifications done in the SUT.}
    \label{tab:toy_versions}
    \begin{tabularx}{\columnwidth}{lll}
        \toprule
        \textbf{Version} & \textbf{Modifications} & \textbf{Location} \\
        \hline
        $v_1$   & \texttt{month + 1} & \texttt{MyDate-Line 9}\\
        $v_2$   & \texttt{year  + 1} & \texttt{MyDate-Line 9}\\
        $v_3$   & \texttt{day   + 1} & \texttt{MyDate-Line 9}\\
        $v_4$   & \texttt{month + 1, year + 1} & \texttt{MyDate-Line 9}\\
        $v_5$   & \texttt{day   + 1, year + 1} & \texttt{MyDate-Line 9}\\
        \bottomrule
    \end{tabularx}
\end{table}

\begin{figure}
    \centering
    \includegraphics[width=\columnwidth]{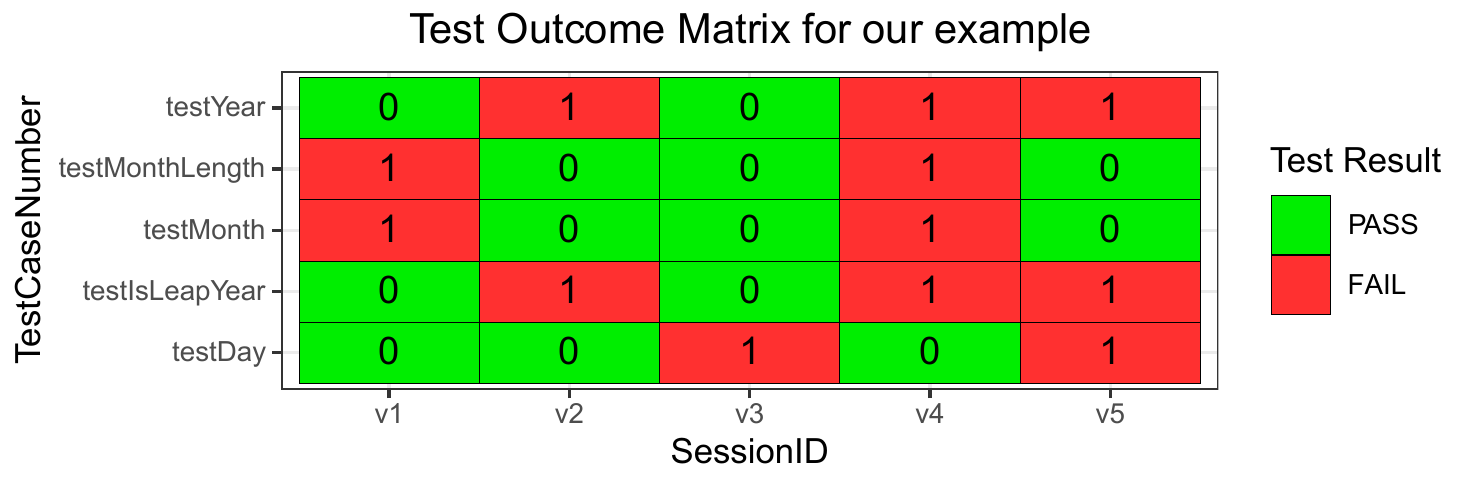}
    \caption{A visual representation of a TOM, where 0 represents that a test passed, whereas 1 represents that a test fail.}
    \label{fig:toy_tom}
    \vspace{-0.5cm}
\end{figure}

For our example, the TOM (Figure~\ref{fig:toy_tom}) reveals some patterns between its rows and columns. Particularly, the rows for \texttt{testMonth} and \texttt{testMonthLength} have the same outcome throughout all five different versions of the SUT. In turn, \texttt{testYear} and \texttt{testLeapYear} also have the same outcome between one another, but are different if compared with the former pair of tests. Even though, from a black-box perspective, the unit tests themselves are independent from one another (i.e., they call different methods of our SUT), they expose some behaviour patterns after introducing different mutants on the SUT. For instance, whenever a modification is performed in the month variable ($v_1$ and $v_4$), both \texttt{testMonth} and \texttt{testMonthLength} fail together (similar to changing the year variable for \texttt{testYear} and \texttt{testLeapYear}). Furthermore, when changing both month and year variables ($v_4$), all those four tests fail together, which in contrast is not seen when changing only year and day ($v_5$), where \texttt{testDay} fails together with \texttt{testYear} and \texttt{testLeapyear}. 

In summary, tests exercising (dis-)similar parts of the SUT where mutants (faults) are introduced are likely to show the same result (pass\slash fail) throughout cycles of test execution. Therefore, each mutant needs to be isolated into different versions of the SUT, as opposed to a single SUT version with several mutants. Otherwise, we lose granularity of the test behaviours for different types of faults since we may not be able to distinguish the corresponding mutants killed by the different pairs of test cases.

There are still many factors that can affect how the TOM is representative of the tests' behaviour. For instance, tests often have more than just two possible outcomes (e.g., interrupted, blocked, not executed), particularly at integration and system level where system's dependencies and issues also affect test execution (e.g., service availability, or performance bottlenecks). That and many other challenges have been discussed in literature when test history is used for test prioritisation~\cite{Kim2002_failrate,Zhu2018_cofail,Haghighatkhah2018_jss}, such as the reliability of the test execution themselves~\cite{Zhu2018_cofail}. Through mutation testing, we are able to execute all tests on all SUT version, and thus can avoid the many sources of noise that history-based diversity calculation suffer from.

Nonetheless, our approach also introduces risks, such as the inaccuracies with mutation operators or the performance bottleneck often seen in mutation testing~\cite{Kintis2018_pit}. Nonetheless, this paper focuses on proposing and evaluating the use of mutation testing in behavioural test diversity scenarios. For now, we thus consider only passing and failing tests, while removing inconsistencies within and across columns of the TOM. At this stage, we do not tailor the approach to use and verify different types of mutation operators or subsumption relations between mutants~\cite{kurtz2014mutant}. Future work can further investigate how behavioural diversity and specific types of mutants relate.

\subsection{Calculating distances}

In order to quantify diversity we calculate the distance between the rows of the TOM (i.e., test cases). The goal is to capture how much the tests (rows) differ in terms of their outcome with respect to faults\slash mutants in the SUT (columns). Therefore, for each pair of tests, we consider a confusion matrix for the outcomes of both tests (Table~\ref{tab:conf_matrix}).

\begin{table}
    \centering
    \caption{Confusion matrix comparing the outcomes of test cases for true positive and negatives (TP and TN) or false positives and negative (FP and FN).}
    \label{tab:conf_matrix}
    \begin{tabular}{c|c|c|c}
        \toprule
        \multicolumn{2}{c|}{\multirow{2}{*}{\textbf{Test outcome}}}& \multicolumn{2}{c}{\textbf{$t_b$}} \\
        \cmidrule{3-4}
        \multicolumn{2}{c|}{} & \textbf{Fail} &  \textbf{Pass} \\
        \midrule
        \multirow{2}{*}{$t_a$}& \textbf{Fail} & True Positive & False Positive\\
        \cmidrule{2-4}
        & \textbf{Pass} & False Negative & True Negative \\
        \bottomrule
    \end{tabular}
    \vspace{-0.5cm}
\end{table}

In this paper we focus on two simple distance measures calculated from a confusion matrix: Accuracy (Acc) and Matthew's Correlation Coefficient (MCC). Accuracy measures the percentage of tests with the same outcome (either true positives or true negatives). In other words, we assume that a pair of tests have similar behaviour if they frequently pass and fail together.

\begin{equation}
\footnotesize
    d_{acc}(t_a,t_b) = 1 - \frac{TP + TN}{TP + TN + FP + FN}
\end{equation}

The measure $d_{acc}$ is simple and intuitive, but it is naive since it assumes that two tests that often pass together are similar to one another. In practice, the majority of tests cases pass~\cite{Memon2017_google,Zhu2018_cofail}, hence the number of TN for all pairs in a TOM is likely to be much higher than the TP. Therefore, we decided to also use Matthew's correlation coefficient (MCC, or the $\phi$ coefficient) as a distance measure, since it is often used in two-class classification algorithms. Unlike $d_{acc}$, MCC balances both the true positives and true negatives, yielding a value that varies between $-1$ and $+1$, such that: $+1$ represents a perfect match between both tests, whereas $-1$ indicates full disagreement and 0 represents that the matched outcomes are random. Then, we normalize the distance values from $d_{mcc}$.

\begin{equation}
\footnotesize
\begin{split}
    d_{mcc} =& \frac{(TP * TN) - (FP * FN)}{\sqrt{(TP+FP)\times(TP+FN)\times(TN+FP)\times(TN+FN)}}\\
    \\
    d_{mcc} =& \frac{(1 - d_{mcc})}{2} \text{    (to normalize values between 0 and 1.)}
\end{split}
\end{equation}

We illustrate the MCC measure by calculating below its distance matrix for our example. Using the TOM in Figure~\ref{fig:toy_tom}, we calculate the distances between \texttt{testYear} ($t_1$) and \texttt{testMonth} ($t_2$). Considering \texttt{$t_1$ = \{0,1,0,1,1\}}, and \texttt{$t_2$ = \{1,0,0,1,0\}}, we obtain the following classification: \texttt{\{FP,FN,TN,TP,FN\}}. The entire distance matrix for MCC is shown in Table~\ref{tab:toy_matrices}. For this example, the distance matrix for Accuracy is very similar.

\begin{table}
    \centering
    \caption{Distance matrix based on MCC from the TOM in our example.}
    \label{tab:toy_matrices}
    \begin{tabularx}{\columnwidth}{l|YYYYY}
        \toprule
        \textbf{MCC} & \texttt{test Year} & \texttt{test Month Length} & \texttt{test Month} & \texttt{test IsLeap Year} & \texttt{test Day} \\
        \midrule
        \texttt{testYear}        & 0.00 & 0.58 & 0.58 & 0.00 & 0.58 \\
        \texttt{testMonthLength} & 0.58 & 0.00 & 0.00 & 0.58 & 0.83 \\
        \texttt{testMonth}       & 0.58 & 0.00 & 0.00 & 0.58 & 0.83 \\
        \texttt{testIsLeapYear}  & 0.00 & 0.58 & 0.58 & 0.00 & 0.58 \\
        \texttt{testDay}         & 0.58 & 0.83 & 0.83 & 0.58 & 0.00 \\
    \bottomrule
    \end{tabularx}
\end{table}

For instance, regardless of the distance function, the distance between \texttt{testMonth} and \texttt{testMonthLength} is zero, meaning that both test cases are perceived as identical in terms of their behaviour. Conversely, \texttt{testDay} is the most different of all tests regarding it's behaviour.

\section{Methodology}
\label{sec:methodology}

\begin{figure}
    \centering
    \includegraphics[width=\columnwidth]{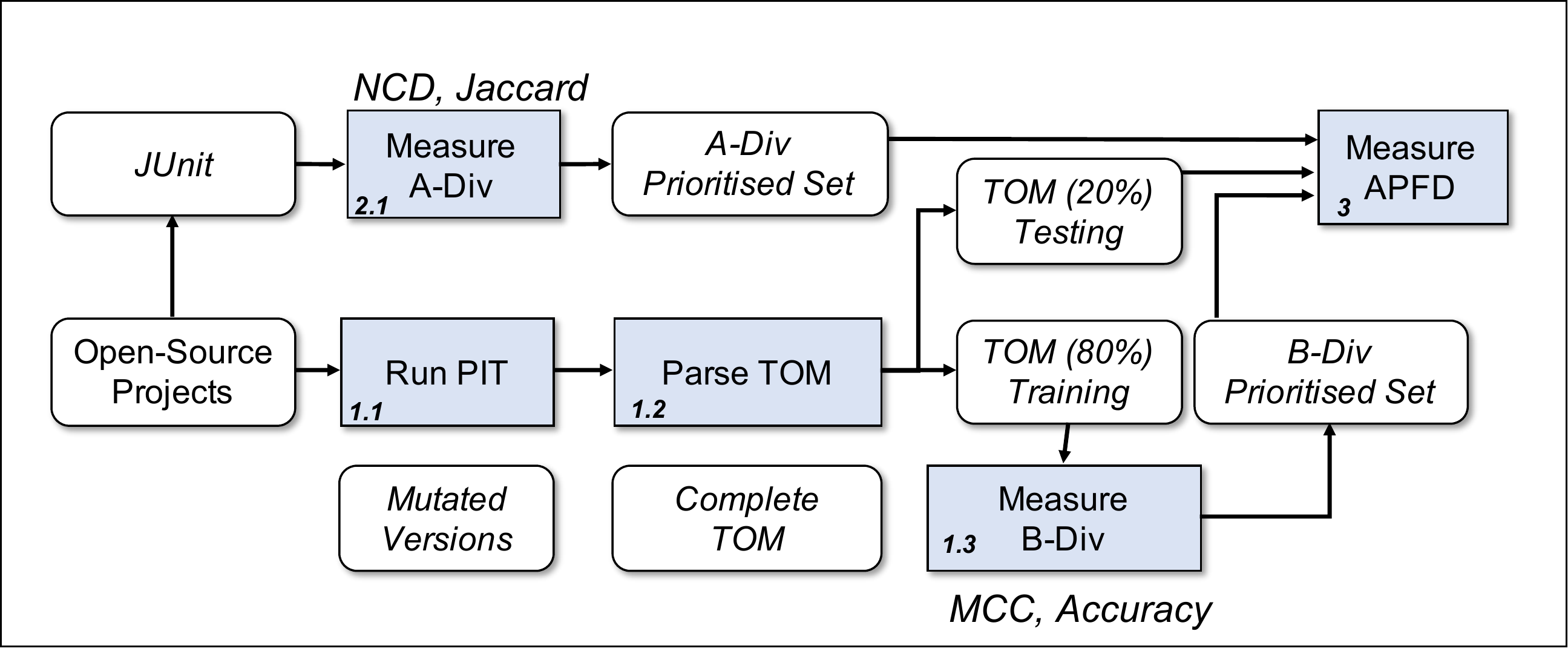}
    \caption{A summary of the steps to execute our experiment. For each open source project, we run PIT to obtain the mutants that either survived or were killed by the tests. This subset of mutants is then used to calculate b-div, whereas the JUnit test code is used to obtain a-div.}
    \label{fig:methodology_study1}
\end{figure}

Figure~\ref{fig:methodology_study1} shows an overview of our methodology. We evaluate our approach using 6 different open-source projects and using the mutation testing tool PIT. Below, we summarize our empirical study as suggested in~\cite{Wohlin2012}: Analyse \textit{behaviour-based diversity}, for the purpose of \textit{evaluation}, with respect to their \textit{effectiveness}, from the point of view of the \textit{tester}, in the context of \textit{test prioritization}.

\textit{Factor and levels:} We use one factor, namely the test prioritisation technique, with four levels: the techniques for measuring behaviour-based diversity (MCC and accuracy), artefact-based diversity, and a Random prioritisation as baseline. For artefact-based diversity we used NCD, since it has shown promising results reported in other empirical studies~\cite{Henard2016,Feldt2016,Haghighatkhah2018_jss}. Consequently, we have a \textit{one-factor with several levels} experimental design and test for statistically significant difference between levels by using a pairwise Mann-Whitney test with Holm-Bonferroni correction~\cite{deOliveiraNeto2019_jss}.

\textit{Dependent variable:} We evaluate the techniques in terms of their APFD (average percentage of fault detection)~\cite{Elbaum2002_apfd}, where faults are represented as mutants in the source code. Using our approach, we instrument PIT to obtain various mutated versions of the SUT and, hence, the behaviour of the test executions. Note, however, that these versions are \textit{not} connected to time or evolution of the SUT, such that there is not a newer or older versions of the SUT.

\textit{Subjects and Objects:} We do not have participants\slash subjects in our experiment. For objects, we choose six open source projects available on github. We searched for small\slash medium Maven Java projects where PIT could be readily connected and quickly executed. The projects are:

\begin{itemize}
    \item Apk-parser: Decodes Android binary xml files, extracting meta information.\footnote{\url{https://github.com/hsiafan/apk-parser}}
    \item Java-string-similarity: Implements string similarity and distance measures.\footnote{\url{https://github.com/tdebatty/java-string-similarity}}
    \item Jfreechart: 2d charting library for Java.\footnote{\url{https://github.com/jfree/jfreechart}}
    \item Lanterna: Library for creating text-based GUI's.\footnote{\url{https://github.com/mabe02/lanterna}}
    \item mp3agic: Library for creating and manipulating mp3 files.\footnote{\url{https://github.com/mpatric/mp3agic}}
    \item sorting-algorithms: Implements a number of sorting algorithms, among them bubble and quick-sort.\footnote{\url{https://github.com/murraco/sorting-algorithms}}
\end{itemize}

Generally, the steps to execute the experiment are: to run PIT and collect the mutation data in each project, extract the TOM, calculate distance matrices using all techniques, prioritise the tests, and collect APFD results. More details on the use of PIT follows in further below under instrumentation. The SUT mutation has two purposes in our study: i) measure b-div, and ii) the mutants killed are used to measure APFD. To avoid validity threats, we divide each TOM into two parts (80\slash 20) used for, respectively, training of the behavioural diversity (purpose `i') and testing the effectiveness of the prioritised subsets (purpose `ii').

\textit{Instrumentation and Operation:}
The PIT mutation testing system in it's default setting executes tests until a first mutant killing occurrence. This results in practice in non-executed tests. For reasons of completeness, we therefore created a plugin for PIT that made sure all test-case, mutant combinations are exercised, with the potential outcome of a KILL (test failure), a SURVIVE (test success), or a time-out.

To measure diversity, we use the open source package MultiDistances\footnote{\url{https://github.com/robertfeldt/MultiDistances.jl}} implemented in Julia\footnote{\url{https://julialang.org/}}. The package offers a variety of artefact-based diversity measures (e.g., Jaccard, Levenshtein and NCD with various compressors), and several ranking\slash prioritization methods. Note that the MultiDistances package does not include the behavioural diversity techniques. Particularly, we use MultiDistances to (i) calculate the a-div of the JUnit test cases from each open-source project, and (ii) rank the tests based on the \textit{maximum mean} of the distance values. The ranking algorithm begins by selecting the pair of tests further apart (i.e., highest distance value), and then, at each step, it adds the test case that has the maximum mean distance to the already selected test cases.

\textit{Controlled variables:} In order to enable comparison between the different approaches, we control the prioritisation budget. We gradually increase the percentage of executed tests from $budget = \{1,5,10,20,\ldots,100\}$ by a factor of 10\%. No in-depth analysis of mutated code coverage in the test runs was done, which is why mutants never resulting in failures, as well as test executions with time-outs, were excluded from the analysis. We used PIT with all mutant operators available which created a vast number of mutants. Note also that a-div does not vary since the information used to measure the distance (i.e., JUnit test cases), does not change between the mutated versions. Moreover, we control for randomness by executing each technique 20 times, and analysing it based on the mean and a 95\% confidence interval.\\

\section{Results and Analysis}
\label{sec:results}

We executed all techniques on the 6 different open-source projects and here present general insights as well as the results of the test executions. Table \ref{tab:mutation} summarizes the information about the mutation process. Execution times span from minutes for small projects to days for the largest project in the investigation. The time of execution seems more strongly correlated to the number of killed mutations rather than generated mutants, as illustrated by the large difference in generated mutants but small difference in execution time for the java-string-similarity and lanterna projects.

\begin{table}
    \centering
    \footnotesize
    \caption{Project mutation statistics. Execution time is shown in hours.}
    \label{tab:mutation}
    \begin{tabularx}{\columnwidth}{lXXXXX}
    \toprule
         \textbf{Project} & \textbf{Classes} & \textbf{TC} & \textbf{Gen.} & \textbf{Killed} & \textbf{Exec} \\
 &  &  & \textbf{Mutants} & \textbf{Percent}& \textbf{time} \\
        \midrule
        jfreechart &637&2177& 271284 & 4\% & 30:22\\
        mp3agic &37&476& 22526 & 14\% & 6:13\\
        apk-parser &125&11& 19331 & 23\% & 4:26\\
        java-string-similarity &26&23& 6481 & 74\% & 2:13\\
        lanterna &202&35& 75519 & 4\% & 1:46\\
        sorting-algorithm &7&7& 776 & 64\% & 0:14\\
        \bottomrule
    \end{tabularx}
\end{table}

\begin{figure*}
    \centering
    \includegraphics[width=0.9\textwidth]{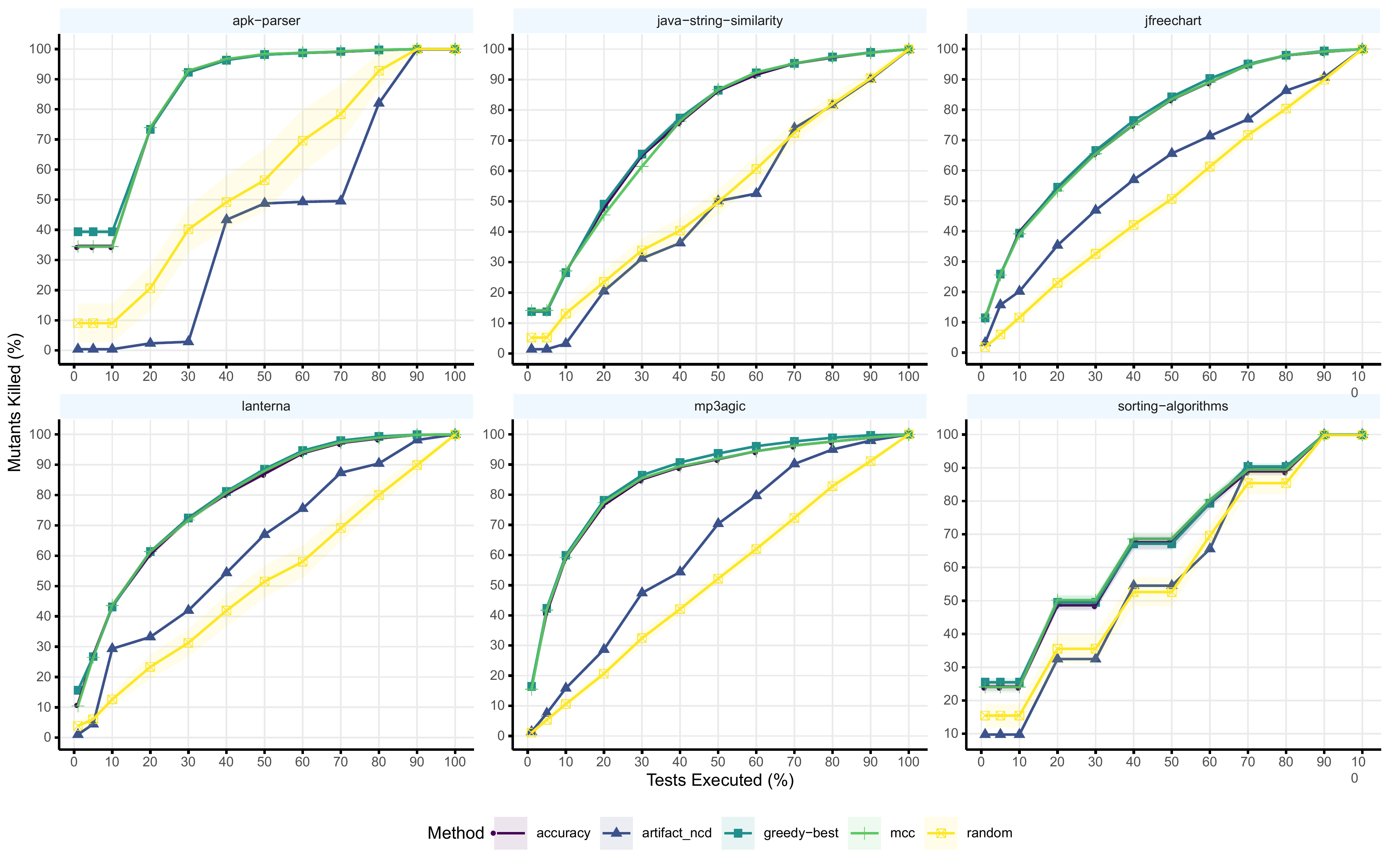}
    \caption{The graphs show the percentage of mutants killed for budget values spanning from 1 to 100\% (in percent of total available test cases).}
    \label{fig:results_curves}
\end{figure*}

The charts in Figure~\ref{fig:results_curves} show how each technique performs in terms of mutants killed for growing subset size. Recall that we evaluate APFD using only 20\% of the set of killed mutants from Table~\ref{tab:mutation}. For instance, if we consider jfreechart, a total of 10851 mutants were killed by its test suite (4\% kill rate), such that we use 8681 of those mutants (80\%) to obtain the behavioural diversity and the remaining 2170 (20\%) to evaluate APFD. In other words, a prioritisation technique with 100\% mutant coverage for jfreechart means that the prioritised test suite is able to reveal those remaining 2170 mutants.

As a reference for comparison, we implemented a greedy one-step look-ahead approach that optimises based on absolute fault knowledge of all test-cases. It minimizes the number of remaining faults, leading to near-optimal prioritisation for all budgets. That way, we see how far each prioritisation-technique is from near-optimal prioritisation. An optimal prioritisation is NP-hard to compute, and would either require prohibitive exhaustive search or backtracking, which would not have been scalable for the problem at hand.

The test prioritization using behavioral metrics converges fastest across the board, and even on par with the greedy-best algorithm that acts as an approximation of the best possible prioritization. The choice of behavioral metric though does not seem to matter for any of the investigated projects. Note that for the smaller projects, namely \texttt{java-string-similarity}, \texttt{sorting-algorithms} and \texttt{apk-parser}, random prioritisation outperform the artefact-based diversity measures. Recall that the input used to measure a-div information are the code in each JUnit test method. Looking closely at the data, we identified that the few and small test methods did not provide much information to the NCD measure.

In turn, when considering the bigger projects (mp3agic and jfreechart), \texttt{a-div} performed reasonably better than random, being able to reveal between 5\% to 15\% more faults. Nonetheless, \texttt{b-div} was even more beneficial for those projects. For instance, considering a 30\% budget, \texttt{b-div} revealed an average between 19\% and 31\% more mutants when compared to \texttt{a-div}. Details of these differences, for the budgets of 10\% and 30\% in two projects are presented in Table~\ref{tab:descriptive_stats}.

\begin{table*}
    \caption{Median percentage of mutants killed and corresponding standard deviation (SD) for each project in our dataset. The highest percentage for each project is highlighted. Recall that greedy-best is not a level in our factor and simply acts as a relative comparison between a higher bar of mutants that can be revealed for that budget.}
    \label{tab:descriptive_stats}
    \centering
    \begin{tabularx}{\textwidth}{llXXXXXXXXXXXX}
    \toprule
    &  & \multicolumn{2}{l}{\textbf{apk-parser}} & \multicolumn{2}{l}{\textbf{java-string-similarity}} & \multicolumn{2}{l}{\textbf{jfreechart}} & \multicolumn{2}{l}{\textbf{lanterna}} & \multicolumn{2}{l}{\textbf{mp3agic}} & \multicolumn{2}{l}{\textbf{sorting-algorithms}} \\
    \cmidrule(lr){3-4} \cmidrule(lr){5-6} \cmidrule(lr){7-8} \cmidrule(lr){9-10} \cmidrule(lr){11-12} \cmidrule(lr){13-14}
    (\%) & Technique & Median & SD & Median & SD & Median & SD & Median & SD & Median & SD & Median & SD \\
    \midrule
    & greedy-best & 39 & 1.56 & 26.57 & 1.52 & 39.34 & 1.22 & 43.58 & 2.02 & 59.98 & 1.01 & 25.65 & 2.53\\
    & accuracy & 34.28 & 1.54 & 26.88 & 1.65 & \cellcolor{blue!15}39.95 & 1.16 & 43.08 & 2.36 & 58.98 & 1.27 & \cellcolor{blue!15}24.03 & 2.78\\
 10 & mcc & \cellcolor{blue!15}34.8 & 1.64 & \cellcolor{blue!15}27.38 & 1.76 & 39.11 & 1.29 & \cellcolor{blue!15}43.58 & 1.97 & \cellcolor{blue!15}59.15 & 1.19 & 23.7 & 3.64\\
    & artifact-ncd & 0.39 & -- & 3.25 & -- & 20.17 & -- & 29.33 & -- & 15.83 & -- & 9.74 & --\\
    & random & 0.78 & 13.88 & 14.81 & 5.96 & 11.55 & 2.95 & 11.91 & 4.3 & 10.19 & 3.37 & 12.01 & 7.62\\
\midrule
    & greedy-best & 92.3 & 0.86 & 65.92 & 2.11 & 66.86 & 1.31 & 72.51 & 1.57 & 86.57 & 0.72 & 49.35 & 4.22\\
    & accuracy & 92.76 & 1.03 & \cellcolor{blue!15}65.31 & 1.93 & 65.32 & 1.35 & \cellcolor{blue!15}72.51 & 2.17 & 85.22 & 0.84 & 47.4 & 3.36\\
 30 & mcc & \cellcolor{blue!15}92.82 & 0.88 & 61.46 & 1.63 & \cellcolor{blue!15}65.74 & 1.15 & 72.1 & 1.81 & \cellcolor{blue!15}85.41 & 0.67 & \cellcolor{blue!15}49.35 & 3.42\\
    & artifact-ncd & 2.85 & -- & 31.24 & -- & 46.88 & -- & 41.96 & -- & 47.43 & -- & 32.47 & --\\
    & random & 40.43 & 16.41 & 32.56 & 8.55 & 33.95 & 5.65 & 29.53 & 9.71 & 31.73 & 4.97 & 38.64 & 9.41\\
 \midrule
    & greedy-best & 98.06 & 0.5 & 86.71 & 1.49 & 84.41 & 1.01 & 88.9 & 1.16 & 93.8 & 0.47 & 67.86 & 4.28\\
    & accuracy & \cellcolor{blue!15}98.19 & 0.29 & 86.21 & 1.27 & 83.41 & 0.91 & 87.27 & 1.66 & \cellcolor{blue!15}91.97 & 0.67 & 66.88 & 4.31\\
 50 & mcc & \cellcolor{blue!15}98.19 & 0.5 & \cellcolor{blue!15}86.41 & 1.39 & \cellcolor{blue!15}83.68 & 0.98 & \cellcolor{blue!15}87.98 & 1.57 & 91.87 & 0.54 & \cellcolor{blue!15}68.18 & 3.96\\
    & artifact-ncd & 48.77 & -- & 50.1 & -- & 65.59 & -- & 67.01 & -- & 70.39 & -- & 54.55 & --\\
    & random & 57.5 & 22.18 & 50.1 & 9.19 & 50.35 & 4.1 & 50.81 & 11.17 & 52.21 & 3.41 & 51.95 & 9.38\\
\bottomrule
    \end{tabularx}
    \vspace{-0.5cm}
\end{table*}

We performed a Mann-Whitney U test to test for statistically significant differences between mcc, accuracy and random. Since the artefact-ncd technique did not have a variance, the comparison of median from Table~\ref{tab:descriptive_stats} already shows a clear difference between the artefact-based diversity and the other techniques. For all pairwise comparisons, we did not detect a statistically significant difference between mcc and accuracy, which confirms the similar curves shown in Figure~\ref{fig:results_curves}. Moreover, both mcc and accuracy were significantly different from random at $p < 0.001$ for all investigated budgets. In short, both our visual and statistical analysis confirm that behavioural diversity consistently outperforms the other techniques.

Note, however, that the projects have disparate number of test cases covering its classes (Table~\ref{tab:mutation}). For instance, apk-parser has only 11 test cases for its 125 classes, such that many mutants are likely to survive. Similarly, the three larger projects have the smallest percentage of mutants killed, and yet, very high APFD when using b-div. On one hand, the analysed projects require more and better tests which can increase mutation kill rates. On the other hand, the prioritisation is orthogonal to the mutation scores from Table~\ref{tab:mutation}, such that b-div outperforms the other techniques regardless of the percentage of mutants killed. For instance, results for b-div are still close to greedy for both java-string-similarity and jfreechart, despite the low mutant kill rate from PIT (respectively, 74\% and 4\% of mutants killed from Table~\ref{tab:mutation}).

\section{Discussion}
\label{sec:discussion}

Our results show that behavioural diversity in combination with mutation testing significantly kills more mutants when compared to typical artefact-based diversity and random prioritisation. Below we discuss these findings in connection to our research questions.\\

\noindent \textit{RQ1: How can we represent and capture behaviour diversity?}

We represent behavioural diversity through test executions, captured here in the Test Outcome Matrix (TOM). Then, we measure the distance between those tests by comparing their paired outcomes (e.g., pass\slash fail) throughout the various mutations. We used two distance measures, i.e., Accuracy and MCC, to compare rows of the TOM and obtain a distance matrices with distances between all pairs of test cases. Other prioritisation techniques that combine diversity with execution information do not measure distance on execution outcomes and, instead, measure distances on, e.g., code artefacts~\cite{Haghighatkhah2018_jss}, or the inputs, execution traces, and\slash or outputs of the tests~\cite{Feldt2008,Feldt2016}.\\

\noindent \textit{RQ2: How can we use mutation testing to calculate behavioural diversity?}

Prioritisation techniques that rely on the history of test failures report improved effectiveness~\cite{Kim2002_failrate,Zhu2018_cofail, Haghighatkhah2018_jss}. But researchers also report on the challenges of failure data collection, such as low confidence in the dataset due to very few test failures, skewed data~\cite{Zhu2018_cofail}, or even the volume of data being insufficient to achieve reliable results~\cite{Haghighatkhah2018_jss,Zhu2018_cofail}. Here, we use mutation testing to overcome those challenges by systematically creating mutated versions of the SUT.

We also noted in our results that mutants, created by different mutation operators, are more often killed than others. For instance, out of the in total almost 30,000 mutant kills throughout the projects, mutants based on the \texttt{InvertNegsMutator} were killed only four times, whereas \texttt{RemoveConditionalMutator} and \texttt{NonVoidMethodCallMutator} mutants were killed more than 2000 times each. For a more elaborate understanding of operator suitability, the total number of mutants of each type being produced must be studied and incorporated. Suitability may further vary between projects of different purpose. Future work can also explore questions regarding mutation operator hierarchies and subsumption for the different projects analysed and their connection to b-div~\cite{kurtz2014mutant}.

Furthermore, mutation testing introduces other challenges to our approach. A mutation tool must be available in the programming language of choice, and it has to perform well because of the sheer amount of mutants being tested. For mutation testing and creation of the TOM, this problem is of quadratic complexity in the number of test-cases (rows) and mutants created (columns). For instance, we've selected PIT over MuJava as the former mutates over byte-code instead of Java source code, making it faster and independent of the availability of source code, at least in principle. However, PIT halts test execution upon finding the first test that kills a specific mutant, whereas our approach requires the execution to continue in order to see the outcome of each test for each mutant created. This configuration was not a simple parameter choice in PIT, hence we had to create a plugin.\\

\noindent \textit{RQ3: How do artefact and behaviour-based diversity differ in terms of effectiveness?}

In our experiment, b-div is able to reveal substantially more faults (mutants) than a-div measures in all projects investigated. For projects with more tests, b-div performs better regardless of the distance measure used (Accuracy or MCC). Considering jfreechart, b-div was able to reveal 85\% of the killed mutants after executing only 30\% of the test suite. On one hand, some projects have more test cases than others, which affects artefact diversity since it uses the test cases themselves to measure distance. For the projects with fewer test cases (apk-parser and sorting-algorithms), a-div performed equal or worse than random, such that the projects had low test coverage to begin with, whereas a-div performs better on the projects with more test cases.

Additionally, the distances are measured on small JUnit test methods, which may not contain meaningful information for the NCD measure to properly capture the distinction between tests~\cite{deOliveiraNeto2018_astCI}. Those results for a-div are not surprising, since the diversity is measured based on the content of the test specification\slash script, as opposed to its execution. In short, our results show that behavioural diversity is more effective in revealing mutants than artefact diversity and is independent of the content of the test case. Conversely, b-div requires failure information which, even through mutation testing, can be costly to obtain for larger projects.

With respect to the prioritisation algorithm, artefact-based and behaviour-based diversity can be used interchangeably since the diversity ranking algorithm operates on a distance matrix, which can be generated by both approaches. Furthermore, behavioural diversity does not require availability of a test specification (e.g., script, requirements, code) since it only requires information about the tests' outcomes.
However, future work should consider also set-based diversity maximization such as used in ~\cite{Feldt2016,shin2018theoretical}.
\\

\noindent \textit{RQ4: What are the trade-offs in applying behavioural diversity in test prioritisation?}

Similar to related work, we observe an increased effectiveness of diversity-based prioritisation that consider failure information~\cite{Kim2002_failrate,Haghighatkhah2018_jss,Zhu2018_cofail}. However, other approaches aim to identify how likely tests are to \textit{fail again}, based on past failure information. Behavioural diversity, instead, identifies tests that are likely to \textit{fail together} for a variety of mutants, such that those tests would reveal the same fault. The use of mutation testing mitigates the risks with inconsistent failure information from different builds, such as tests that could not be executed (i.e., unknown outcome), or those that are flaky. Since the main goal with test prioritisation is to reveal the different faults as early as possible, behavioural diversity is a strong candidate because tests that consistently fail together due to the same mutant would be measured as similar, such that only one of them would be ranked with higher priority. Our approach can also be used from scratch while history-based approaches would need to wait until sufficient number of tests have been performed, over time. Future works should consider how to combine history- and mutation-based behavioral diversity.

When relying on mutation testing our prioritisation technique may not be applicable for manual test cases since it is infeasible to run them for many mutants. This can be a relative benefit of artefact diversity that have been used on various types of tests (manual~\cite{deOliveiraNeto2018_apsec}, automated~\cite{deOliveiraNeto2018_astCI}, system-level~\cite{Cartaxo2011}, etc.). Thus, the approaches can complement each other and different combinations can be considered in different contexts. In short, if failure information from historical test logs are available or mutation tools applicable, behavioural diversity should be considered for test prioritisation, in isolation or as a complement to artefact diversity.

\subsection{Limitations and Threats to validity}

One \textit{construct validity} threat of our evaluation is the choice of mutation tool and corresponding mutation strategy. PIT has been recommended in literature and used in various studies~\cite{Kintis2018_pit}, mainly because of its efficiency and the set of mutation operators offered (\url{http://pitest.org/quickstart/mutators/}). Since it operates on the Bytecode level, PIT reduces costs in changing and compiling source code. Nonetheless, we aim to investigate the effect that different mutation operators and tools have on diversity (both artefact and behavioural) in future studies. Furthermore, our experimental study is limited on its usage of artefact-based diversity. The use of NCD in small JUnit test cases affects artefact diversity, since the various short files do not carry much information. In ~\cite{deOliveiraNeto2018_apsec} and \cite{deOliveiraNeto2018_astCI}, authors also discuss this issue in empirical studies where NCD and Jaccard are used on short textual data (test requirements) and have lower test coverage when compared to the same a-div measures on larger test specification (e.g., XML files with system test steps).

Additionally, we evaluate APFD using a population of killed mutants instead of real faults. In this paper we do not investigate the types of the killed mutants and the similarity among them. Constellations of very similar mutants could, potentially, favor the b-div techniques. Conversely, the split is done at random points of the sample and each analysis is repeated 20 times, such that those correlations would lead to higher standard deviation in our APFD for each technique, which is not the case.

Our \textit{internal validity} threats are mainly associated to the execution of our experiment, such as collecting data for which mutants were killed, making sure that all test cases are executed against all generated mutation version, or removing data that is not needed in our approach (e.g., mutants that were not \textit{covered} by any test). The execution of our study involves different tools and frameworks (PIT, the MultiDistances package, R scripts for the analysis) that are are integrated and executed on Docker to mitigate risks with platform specific environments or package dependencies.

In turn, we mitigate \textit{conclusion validity} threats in our statistical analysis by using non-parametric tests and avoid assumptions regarding data distributions~\cite{deOliveiraNeto2019_jss}. Note that most conclusions can be drawn by comparing descriptive statistics due to the clear APFD difference between the techniques. Even though we use a variety of distance measures on a controllable experimental study, our choice of constructs hinders our \textit{external validity}. Future studies include comparison with state-of-the-art prioritisation techniques that also use history information. Nonetheless, the goal of this study is to explore the applicability and trade-off of b-div, whereas future studies can focus on identifying and understanding the limitations between both families of test diversity measure (e.g., when one is preferable, and how they complement each other).

\section{Concluding Remarks}
\label{sec:conclusions}
In this paper we proposed an approach that uses mutation testing tools to capture and calculate behavioral diversity (b-div) of test cases. Unlike artefact-based diversity, our approach uses the patterns of test outcomes to calculate the distance between test cases and the selects test cases with large distances. In particular, we used two distance measures, namely MCC and Accuracy calculated based on confusion matrices. We evaluate our approach by prioritising the test cases of six open-source projects using the PIT mutation testing tool. Our results show that behavioural diversity can kill more mutants on average when compared to artefact-based diversity and random selection. In turn, the use of mutation tools enabled the collection and analysis of failure information which is often a hindrance in applying and evaluating history-based prioritisation techniques~\cite{Zhu2018_cofail,Haghighatkhah2018_jss}.

Future work includes the investigation of the boundaries between artefact and behavioural diversity to better understand when they can complement each other. Furthermore, we intend to investigate the use of mutation tools to complement history-based prioritisation schemes, if our measures can be used to extend the diversity-aware mutation adequacy criterion~\cite{shin2018theoretical}, as well as observing whether these techniques are affected by specific types of mutants~\cite{kurtz2014mutant}.

\bibliographystyle{IEEEtran}
\bibliography{IEEEabrv,IEEEexample}

\end{document}

%% file: javacode_sut.tex
\begin{lstlisting}[caption={Our class under test is a wrapper for LocalDate from Java API.},label={code:example_sut},language=Java]
package sut;
import java.time.LocalDate;

public class MyDate {

	private LocalDate myDate;

	public MyDate(int year, int month, int dayOfMonth) {
		myDate = LocalDate.of(year, month, dayOfMonth);
	}

	public int getYear() {
		return myDate.getYear();
	}

	public int getMonth() {
		return myDate.getMonthValue();
	}

	public int getDayOfMonth() {
		return myDate.getDayOfMonth();
	}

	public int getMonthLength() {
		return myDate.getMonth().maxLength();
	}

	public boolean isLeapYear() {
		return myDate.isLeapYear();
	}
}
\end{lstlisting}

%% file: javacode_test.tex
\begin{lstlisting}[caption={Example of unit tests to cover the SUT used in our example.},label={code:example_test},language=Java]
package test;
import static org.junit.*;
import sut.MyDate;

public class MyDateTest {
	private static int YEAR = 2019;
	private static int MONTH = 10;
	private static int DAY = 12;

	private MyDate date;

	@Before
	public void setUp() {
		date = new MyDate(YEAR, MONTH, DAY);
	}

	@Test
	public void testYear() {
		assertEquals(YEAR, date.getYear());
	}

	@Test
	public void testMonth() {
		assertEquals(MONTH, date.getMonth());
	}

	@Test
	public void testDay() {
		assertEquals(DAY, date.getDayOfMonth());
	}

	@Test
	public void testMonthLength() {
		final int OCT_LENGTH = 31;
		assertEquals(OCT_LENGTH, date.getMonthLength());
	}

	@Test
	public void testIsLeapYear() {
		// 2019 is not a leap year.
		assertFalse(date.isLeapYear());
	}
}
\end{lstlisting}